\documentstyle[12pt]{article}
\renewcommand{\baselinestretch}{1.25}
\newcommand{\be}{\begin{equation}}
\newcommand{\ee}{\end{equation}}
\newcommand{\bee}{\begin{eqnarray}}
\newcommand{\eee}{\end{eqnarray}}
\newcommand{\al}{\alpha}
\begin{document}
\begin{titlepage}
\vspace*{3.cm}
\begin{center}
      {\Large \bf The $\eta N$ S-wave scattering length in one resonance
      model}
\vspace*{10mm}\\
Mijo Batini\'{c},  Alfred \v{S}varc   \\
{\em Rudjer Bo\v{s}kovi\'{c} Institute,  Zagreb,
 \vspace*{1.cm} Croatia } \\
 Zagreb, \today
\end{center}
      We show  that  the  S-wave  $\eta  N$  scattering  length  can  be
      extracted  in  a  model  independent  way  within the scope of the
      multichannel model, but with the restricting assumption that  only
      one  resonance  is included per partial wave.  One has only to use
      the information on $\pi N$ elastic S-wave T-matrix at  the  $\eta$
      production  threshold,  and  near threshold $\pi^{-} p \rightarrow
      \eta n$ total cross section.  The results are independent  of  the
      particular  parametrization  of  the  elastic  T-matrix and of the
      number  of  channels.   For  the  described  model  the  unitarity
      requires  that  the  number  of  channels  must be at least three.
      These assumptions are more general then  the  assumptions  of  the
      existing single resonance models up to now used for extracting the
      $\eta N$ S-wave scattering length.  The agreement of our  approach
      with  former  estimates is anticipated and confirmed for all cases
      for which the input data agree with the commonly accepted values.
\end{titlepage}
\renewcommand{\baselinestretch}{1.8}
\newpage
\setcounter{page}{1}
      \section{Introduction}
      In 1985 Bhalerao and Liu \cite{Bha85} have constructed  a  coupled
      channel  isobar  model  for  the $\pi N \rightarrow \pi N$, $\pi N
      \rightarrow \eta N$ and $\eta N \rightarrow \eta N$ T  -  matrices
      with  $\pi N$, $\eta N$ and $\pi \Delta$ ($\pi \pi N$) as isobars.
      A single  resonance  separable  interaction  model  for  S$_{11}$,
      P$_{11}$, P$_{33}$ and D$_{13}$ partial waves has been used.  They
      have used only $\pi N$ elastic scattering  data  as  a  constraint
      while their prediction for the $\eta$ production cross section has
      been  compared  with,  at  that  time   the   most   recent   data
      \cite{Bro79}.   Their conclusion has been that the S-wave $\eta N$
      interaction is attractive, and they have extracted for the  S-wave
      scattering length the value of $a_{\eta N}=(0.27 + i \, 0.22)$ fm.

      Wilkin  \cite{Wil93}  based his calculation on an S-wave threshold
      enhancement calculation, used the $\eta$ total cross section  near
      threshold  to  fix the imaginary part of the T-matrix and obtained
      the real part by  fitting  the  $\pi^{-}  p  \rightarrow  \eta  n$
      production  cross section up to the center of mass momentum in the
      $\eta n$ system of  1.2  (fm$^{-1}$).   He  quotes  the  value  of
      $a_{\eta N} = (0.55 \pm 0.20 + i\,0.30)$ fm.

      Abaev  and  Nefkens \cite{Aba94} have also used a form of a S-wave
      single resonance  model,  adjusted  the  resonance  parameters  to
      reproduce the $\pi^{-} p \rightarrow \eta n$ production channel to
      the best of their ability  and  extracted  the  S-wave  scattering
      length as:  $a_{\eta N} = (0.62 + i \, 0.30)$ fm.

      Arima  et  al  \cite{Ari92}  have studied the nature of the S-wave
      resonances  S$_{11}$(1535)  and  S$_{11}$(1650)  concerning  their
      couplings with the $\eta N$ channel using the two quark-model wave
      functions with pure intrinsic spin states for  the  isobars.   The
      dynamical coupling of the isobars to $\pi N$ and $\eta N$ channels
      are  described  by  the  meson-quark  coupling.   In  addition  to
      analyzing  the agreement of the model with the $\pi N$ elastic and
      $\eta$ production data they have obtained  the  S-wave  scattering
      length $a_{\eta N} = (0.98 + i\,0.37)$ fm.

      In  this  article we show that the agreement of the listed results
      with the exception of  Arima  et.   al.   \cite{Ari92}  is  to  be
      expected as the knowledge of $\pi N$ elastic T-matrix, originating
      from any of a world collection of PWA
      \cite{Cut79,Hoe83,Ka84,SAID}
      and the threshold value of the  $\pi^{-}  p  \rightarrow  \eta  n$
      cross  section  are  sufficient  to  calculate the $\eta N$ S-wave
      scattering length value in any, single  resonance  model  and  all
      cited  models  indeed  do  assume using only one resonance for the
      dominant S$_{11}$ partial wave.   Variations  among  the  obtained
      $\eta  N$  S-wave  scattering  length  values  can  be  completely
      attributed to the somewhat different assessment of the input data.

\section{Formalism}
      For the convenience of the reader we give the  collection  of  the
      essential, but well known formulae.

      The  model  of  using  only  one resonance per partial wave is the
      simplest, manifestly unitary, way to  construct  the  multichannel
      T-matrix  for any choice of connected two body processes.  For any
      number of two-body channels, T-matrix is given by
\be
      T_\al(W) = \left( T_{ab,\al}(W) \right) \equiv
      \left(
      \begin{array} {cccc}
         T_{11,\al}(W)  & T_{12,\al}(W) & . & T_{NN,\al}(W) \\
         T_{21,\al}(W)  & T_{22,\al}(W) & . & T_{2N,\al}(W) \\
         .              & .             & . & .             \\
         T_{N1,\al}(W)  & T_{N2,\al}(W) & . & T_{NN,\al}(W) \\
      \end{array}
      \right)
\ee
      where  $W$ is total center of mass energy, $a,b = 1,..., N$ denote
      two body channels involving one nucleon, e.g.:
      \bee
      T_{11} &=& T_{\pi  N\, \pi  N} \equiv T_{\pi \pi } \nonumber \\
      T_{22} &=& T_{\eta N\, \eta N} \equiv T_{\eta\eta} \\
      T_{12} &=& T_{\pi  N\, \eta N} \equiv T_{\pi \eta} \nonumber \\
      ...    & & \ \ \ ... \hspace*{1.0cm} ...           \nonumber
      \eee
      and $\al$ denotes all additional quantum numbers (isospin, angular
      momentum, ...).  From here on, we restrict our  analysis  to  only
      one  resonance  (no  background)  per  partial  wave.   With  that
      restriction the following relation among matrix  elements  can  be
      written:
\bee
\label{eq:SRM}
      T_{ab}(W)               &   =  & y_a(W) y_b(W) T_0(W)   \nonumber \\
      \sum_{a=1}^{N} y_a(W)^2 &   =  & 1                      \\
      T_{0}(W)                &\equiv& \sum_{a=1}^N T_{aa}(W) \nonumber
\eee
      where  $y_a(W)$  are  real quantities and are related to resonance
      branching ratios $x_a$ and for Breit-Wigner  parametrisation  $x_a
      \equiv   y_a(M_R)^2$.    Eq.~(\ref{eq:SRM})  gives  relations  for
      diagonal elements:
\bee
\label{eq:SRT1}
      \frac{T_{aa}}{T_{bb}}&=&P_{ab}
      \ \ \ \ P_{ab}\  {\rm is\ real\ positive\ number}
\eee
      and  for non  diagonal ones:
\bee
\label{eq:SRT2}
      T_{ab} & = & \pm \sqrt{T_{aa}T_{bb}}
\eee
      what  means  the  proportionality between any two complex T-matrix
      elements of the multichannel T-matrix.  This fact does not  depend
      on  any particular parametrization of the T-matrix elements either
      in the case of Breit-Wigner parametrization  or  in  the  K-matrix
      approach.   It  only  depends on the single resonance character of
      the model.  We define the unitary multichannel S-matrix:
\bee
      S_{\al}(W) & = &  1 + 2\,i\,T_{\al}(W) \nonumber \\
      S_{\al}^{\dagger}(W)\,S_{\al}(W) & = &
      S_{\al}(W)\,S_{\al}^{\dagger}(W) = 1\,.
\eee
      The partial wave scattering amplitudes are defined as:
\be
      f_{ab,\al}(W) = \frac{T_{ab,\al}(W)}{\sqrt{p_a p_b}}
                                                \ \ \ \ a,b = 1,2,...N
\ee
      where $p_a$ is the center of mass momentum of particles in channel
      $a$
\be
      p_a = \frac{[W^2-(M_a+m_a)]^2 [W^2-(M_a-m_a)^2]}{2 W}
                                                \ \ \ \ a,b = 1,2,...N
\ee
      and $M_a$, $m_a$ are the corresponding masses.

      The total cross sections is given by
\bee
\label{eq:XSf}
      \sigma^{\rm tot}_{a \rightarrow b,\al} (W)
      &=& 4\pi\, \frac{|T_{ab,\al}(W)|^2}{p_a^2} \nonumber
\eee
      and using Eq.~(\ref{eq:SRT2}) we get:
      \bee
      \label{eq:XSCH}
       \sigma^{\rm tot}_{a \rightarrow b,\al} (W)
      & = & 4\pi\, \frac{|T_{aa,\al}(W)|\,|T_{bb,\al}(W)|}{p_a^2}.
      \eee
      The total cross section summed over all additional quantum numbers
      $\alpha$ is given by:
      \bee
\label{eq:XST}
      \sigma^{\rm tot}_{a \rightarrow b} (W)
      &=& \sum_{\al} \sigma^{\rm tot}_{a \rightarrow b,\al} (W)
\eee
      To obtain the scattering length does not depend on the details of
      the  model  used,  only  the  unitarity  and  the single resonance
      character of the model are required.  The S matrix for the angular
      momentum $l$ is written in the following way:
      \be
      S_{aa,l}(W)=\frac{1+i \ k_{a,l}(p_a)}{1-i \ k_{a,l}(p_a)}
                   \ \ ; \ \ k_{a,l}(p_a)=\tan \delta_{a,l}(p_a),
      \ee
      $p_a$ is the channel $a$ center  of  mass  momentum,  $l$  is  the
      angular momentum quantum number, and $\delta_{a,l}$ is the channel
      $a$ partial wave phase shift.  For the diagonal S-matrix  elements
      we  apply  the  low  energy expansion\footnote{Different textbooks
      take opposite sign for the scattering length, see \cite{Gol64}  vs
      \cite{Tay72}.     \vspace*{-0.3cm}    Here    we    follow    Ref.
      \cite{Gol64}.}:  \\
      \be
      p_a^{2l+1} \cot \delta_{aa,l}(p_a) = \frac{1}{a_{a,l}}
                                         + \frac{r_{a,l}}{2}\, p_a^{2}
                                    + O(p_a^{4})
      \ee
      which  defines  the  channel  $a$  scattering length and effective
      range $a_{a,l}$ and $r_{a,l}$ , respectively.  \\ In the terms  of
      the  scattering  length  and  the  effective  range the scattering
      amplitude $f_{aa,l}(p_a)$ is given by
      \be
      f_{aa,l}(p_a)
             \approx  \frac{a_{a,l}\, p_a^{2l}}
       {1 - ia_{a,l}\,p_a^{2l+1} + \frac{1}{2}a_{a,l}\, r_{a,l}\, p_a^{2}}
      \label{eq:sampl}
      \ee
      So the expression for the scattering length  $a_{a,l}$  is  given
      as:
      \be
          a_{a,l} = \lim_{p_a \rightarrow 0}
               \frac{f_{aa,l}(p_a)}
      {p_a^{2l} + i\,f_{aa,l}(p_a)\, p_a^{2l+1} -
              \frac{1}{2}r_{a,l}\, f_{aa,l}(p_a) p_a^{2} }
      \label{eq:a0}
      \ee
      and because of the threshold behavior $f_{aa,l} \sim p_a^{2l}$,
      the scattering lengths can be approximated by
      \be
          a_{a,l} \approx   \frac{f_{aa,l}(p_a)}{p_a^{2l}  }
                           = \frac{T_{aa,l}(W_a)}{p_a^{2l+1}}
      \ee
      for  the very small $p_a$ i.e.  for the energy $W_a$ very close to
      the threshold of channel $a$.  Namely for the S-wave we have
      \be
\label{eq:aapr}
          a_{a,0} \approx         f_{aa,0}(p_a)
                          = \frac{T_{aa,0}(W_a)}{p_a}
      \ee
      This expression for the S-wave scattering length $a_{a,0}$ is
      used through this article.

      If   we   apply  the  single  resonance  formulae  (\ref{eq:SRT1},
      \ref{eq:SRT2})  for  any  of  the  single  resonance  models,  the
      scattering  length  for  the  channel  $a$  is proportional to any
      T-matrix element at threshold $W_a$:
\bee
\label{eq:SRa1}
      \frac{a_{a,\al}}{T_{bb,\al}(W_a)}&=&r_{ab,\al}
      \ \ \ \ r_{ab,\al}\  {\rm is \ a \ real \ positive \ number}, \\
\label{eq:SRa2}
      \frac{a_{a,\al}}{T_{bc,\al}(W_a)}&=&r_{abc,\al}
      \ \ \ \ r_{abc,\al}\ {\rm is \ a \ real \ number}.
\eee
      When  only  one channel is opened, the scattering length is a real
      quantity.  Upon opening inelastic channels, the scattering  length
      becomes  complex  and  its  imaginary part is related to the total
      cross sections via the optical theorem.  For the multichannel case
      \cite{Gol64,Tay72} we have:
\be
      Im\ f_{aa}(p_a,\vartheta=0)
      = \frac{p_a}{4\pi}
      \sum_{x}\sigma^{tot}_{a \rightarrow x}(W)
\label{eq:opt}
\ee
      where  $x$  denotes  all  opened  channels.   The  optical  theorem
      (\ref{eq:opt}) is automatically satisfied if S-matrix is a  unitary
      matrix.

      If  we  know one of the S-wave diagonal matrix elements $T_{aa,0}$
      at energy $W_b$ where the channel  $b$  opens  $(b  \neq  a)$  and
      $\sigma_{a  \rightarrow  b}^{\rm tot} / p_b$ near threshold $W_b$,
      where      S-wave      dominates,       and       using       Eqs.
      (\ref{eq:XST},\ref{eq:aapr}) we get
\be
\label{eq:aSR}
      a_{b,0}=\frac{1}{4\pi} \, \frac{T_{aa,0}(W_b)}{|T_{aa,0}(W_b)|^2} \,
                \frac{\sigma_{a \rightarrow b,0}^{\rm tot}(W_b)}{p_b}
        \approx \frac{1}{4\pi} \, \frac{T_{aa,0}(W_b)}{|T_{aa,0}(W_b)|^2} \,
                \frac{\sigma_{a \rightarrow b}^{\rm tot}(W_b)}{p_b}.
\ee
      This value depends only on the fact that a single resonance model
      is used. A particular  parametrization or a number of
      channels does not play any importance.

      We  restrict  our  analysis to the $I=1/2$ $\pi N$ scattering with
      inelastic channels $\eta N$, $\pi \Delta$, $(\pi\pi)_{\rm  S-wave}
      N$,  etc.   The  optical  theorem  (\ref{eq:opt})  for the $\eta N
      \rightarrow \eta N$ scattering amplitude near $\eta  N$  threshold
      with the assumption of the S-wave domination gives:
\bee
      Im\ f_{\eta \eta,0}(p_\eta,\vartheta=0)
      & \ \stackrel{p_\eta \rightarrow 0}{=} \ &
      \frac{p_\eta}{4\pi}
      \left(  \sigma^{tot}_{\eta N \rightarrow \pi N}(W)
        + \sigma^{tot}_{\eta N \rightarrow \pi \Delta}(W) \right. \nonumber \\
    & \ + \ & \left. \sigma^{tot}_{\eta N \rightarrow (\pi\pi)_{\rm S} N}(W)
         + .....
      \right)
\eee
      Using  Eq.~(\ref{eq:aapr}),  isospin  algebra  and  the detailed
      balance we get the lower bound  for  the  imaginary  part  of  the
      $\eta N$ S-wave scattering length:
\be
      Im\ a_{\eta N,0} \ge \frac{3p_\pi^2}{8\pi}
      \frac{\sigma^{tot}_{\pi^- p \rightarrow \eta n}(W_\eta)}{p_\eta},
\ee
      where  the  $p_\pi$  is the c.m.  momentum of the particles in the
      $\pi N$  channel  at  $\eta  N$  threshold  $W_\eta$.   Using  the
      experimental  value  of  the $\eta$ production total cross section
      near $\eta N$ threshold $W_\eta$:
\bee
      \frac{\sigma^{tot}_{\pi^- p \rightarrow \eta n}(W_\eta)}{p_\eta}
      &=& (21.2 \pm 1.8)\ \mu{\rm b} / {\rm MeV}
\label{eq:XStot}
\eee
      taken from Ref.  {\rm \cite{Bin73}}, we obtain the  constraint  on
      the  imaginary  part  of  the  $\eta  N$  scattering  length based
      exclusively on the optical theorem:
\be
      Im\ a_{\eta N,0} \ge (0.24 \pm 0.02)\ {\rm fm}.
\label{eq:aEst}
\ee
      Knowing $\pi N$ elastic T matrix and $\pi^- p \rightarrow \eta  N$
      near $\eta N$ threshold total cross section Eq.~(\ref{eq:aSR}) and
      some isospin algebra gives:
\be
\label{eq:aetan}
      a_{\eta N,0} \approx
         \frac{3}{2}
      \, \frac{p_{\pi}^2}{4 \pi}
      \, \frac{T_{\pi \,\pi }(W_\eta)}{|T_{\pi \,\pi }(W_\eta)|^2}
 \frac{\sigma^{tot}_{\pi^{-} p \rightarrow \eta n}(W_\eta)}{p_{\eta}}\,.
\ee
      The result is independent on the details of  the  model,  on  the
      T-matrix  parametrization,  the number of channels, which, because
      of unitarity, can not be lower then three, etc.

      There  are  several  $\pi  N$   elastic   phase   shift   analyses
      \cite{Cut79,Hoe83,Ka84,SAID}.   The  $\pi^-  p \rightarrow \eta n$
      experimental total cross section  is  as  well  quite  well  known
      \cite{Bin73}.   Using  the  different  $\pi N$ elastic phase shift
      analyses and $\pi^- p \rightarrow \eta n$ total cross section from
      Eq.~(\ref{eq:XStot})  as  input  we get the distribution of values
      for the $\eta N$  scattering  length,  for  the  single  resonance
      model, and analyze its sensitivity to the input.
\section{Numerical results and conclusions}
      In  Tables~1--3 and Fig.~1 we show the allowed spread in the $\eta
      N$ S-wave scattering length within the  framework  of  any  single
      resonance  model,  based on the afore described, model independent
      approach.  Numerical values for the  $\eta  N$  S-wave  scattering
      length,    given    in    Table~1,   were   obtained   using   the
      Eq.~(\ref{eq:aetan}) and the $\pi N$ elastic T-matrices near $\eta
      N$  threshold  ($W=1487$  MeV,  $p_\pi=432$  MeV/c)  from  various
      obtainable phase shift analyses  \cite{Cut79,Hoe83,Ka84,SAID}  and
      the     $\eta$     production    total    cross    section    from
      Eq.~(\ref{eq:XStot})  \cite{Bin73}  .    If   we   introduce   the
      uncertainty  of  0.01  for  real and imaginary part of the input T
      matrix and the  uncertainty  of  1.8  $\mu$b/MeV  for  the  $\eta$
      production  total  cross section the resulting uncertainty for the
      $\eta N$ scattering length is of the order of 0.03 fm for real and
      for imaginary part.  Main contribution to the total uncertainty is
      coming  from  the  uncertainty  of  the  $\eta$  production  cross
      section.  The contribution from the T-matrix uncertainty is $\sim$
      5 \%.  In Table~2 we give the results for the $\eta N$  scattering
      length  given  elsewhere together with the input:  $\pi N$ elastic
      T-matrices and the near threshold $\eta$  production  total  cross
      section  when  possible.  A compilation of all results is shown in
      Fig.~1.  Open circles represent the  $\eta  N$  S-wave  scattering
      length  values obtained on the basis of $\pi N$ elastic T-matrices
      of  references   \cite{Cut79,Hoe83,Ka84,SAID}   and   the   $\eta$
      production  total  cross  section  given  in Eq.~(\ref{eq:XStot}).
      Full normal and  inverse  triangles  represent  the  original  and
      modified  scattering  length values of Ref.  \cite{Bha85}, and the
      full square gives the value given by Ref.  \cite{Aba94}.  The full
      circles  represent  the  value  of Ref.~\cite{Wil93}.  Open square
      represent the  value  obtained  in  our  full  model  \cite{Bat94}
      restricted  to  the  single  resonance only and an open normal and
      inverse triangles represent the values of our full, multiresonance
      model  without  any  restrictions  whatsoever  using three or four
      poles in the P$_{11}$ partial wave.  The full star represents  the
      value  of Ref.~\cite{Ari92}, but let us mention that this value is
      {\em not} obtained within the framework of  the  single  resonance
      model.

      The  Eqs.~(\ref{eq:SRa1}, \ref{eq:SRa2}) can be used as a check if
      some particular model is a single resonance one, or there is  some
      influence  of  higher  resonances  and/or background terms.  If we
      specify  Eqs.~(\ref{eq:SRa1}, \ref{eq:SRa2})  for   all   possible
      variations of our model \cite{Bat94} we get:
\be
  r_{\pi \,\eta ,0} =\frac{a_{\eta N,0}}{T_{\pi \,\pi ,0}(W_\eta)} .
\ee
      For single resonance models the ratio
\be
\label{eq:SRR}
      R = \frac{ Im\ r_{\pi \,\eta ,0} }{  Re\ r_{\pi \,\eta ,0} }
\ee
      must be zero\footnote{ The Im $r_{\pi \, \eta ,0}$ is of course 0,
      but  for  the   reason   of   comparison,   the   scaling   factor
      \vspace*{-0.3cm}  Re  $r_{\pi  \, \eta ,0}$ is introduced.}.  This
      ratio is shown in Table~3 for for all cases where  both  $\eta  N$
      S-wave  scattering length and $\pi N$ elastic T matrix at $\eta N$
      threshold are available and well known.  As we can see this  ratio
      is  very  close  to  zero for any single resonance model.  Certain
      deviations from the exact zero occur because some T  matrices  are
      just  read  off the figures in different publications because they
      have not been available in the numerical  form.   For  the  values
      originating from Table~1 this ratio is of course exactly zero.  \\
      Results are very indicative: \\
      As  it  has  been  shown  in  \cite{Bat95},  the  $\eta  N$ S-wave
      scattering length obtained in the framework  of  the  full  model,
      when  other  resonances  and the background term in particular are
      explicitly introduced in the unitary  way  is  significantly  more
      attractive  then any of the predictions obtained within the limits
      of the single resonance models.  All single resonance models  with
      good  $\pi  N$ elastic T-matrix at threshold lie within the limits
      which are obtained  using  realistic  $\eta  N$  production  cross
      section,  and the $\pi N$ elastic T-matrix.  The solution of Arima
      et al.  \cite{Ari92} is much closer to our  final  value,  and  as
      their   model  is  {\em  not}  of  single  resonance  nature,  and
      henceforth it is not  in  the  contradiction  with  the  suggested
      analyses.   However,  the  prediction  of  Ref.   \cite{Aba94}  we
      consider as non-reliable because the threshold value of  the  $\pi
      N$  elastic T-matrix differs significantly from worldwide accepted
      values.  \\
      {\bf Conclusion:} \\
      {\em The simple, model independent mechanism  for  extracting  the
      $\eta  N$  S-wave scattering length from the near threshold values
      of the $\pi N$ elastic T-matrices and the $\eta $ production total
      cross  section  exists  if  only one resonance is used per partial
      wave.  All reported values, based on the single  S-wave  resonance
      assumption  agree  with our model independent results.  The simple
      criterion is offered which, on the bases  of  the  input  $\pi  N$
      elastic  S-wave  T-matrix  and  the  obtained  $\eta N$ scattering
      length,  gives  the  estimate  of  the  importance  of  non-single
      resonance  ingredients  for  the  presented  model.  The realistic
      $\eta N$ S-wave scattering length based on the full, three coupled
      channel,  multiresonance and manifestly unitary model \cite{Bat94}
      is, and  should  be  very  different  from  any  single  resonance
      prediction,  because  the  coupled  channel  process  can  not  be
      reliably described using such a simplified model.  The analysis of
      \cite{Bat95}  gives  the  new  prediction  for the $\eta N$ S-wave
      scattering  length  with  much  more  attraction  than  previously
      reported.}

\newpage \noindent
\section*{Tables}
\begin{table}[h]
\label{tab:1}
\begin{center}
\begin{tabular}{ccc}
\hline\hline
                       & $T_{\pi N\,\pi N}(W_\eta$)   & $a_{\eta N}$ [fm]  \\
\hline\hline
solution CMB  \cite{Cut79}&\ \ $(0.376+i0.439)$\ \ &\ \ $(0.269+i0.315)$\ \ \\
\hline
solution KH80 \cite{Hoe83}&    $(0.332+i0.393)$    &    $(0.301+i0.356)$ \\
\hline
solution KA84 \cite{Ka84} &    $(0.390+i0.374)$    &    $(0.320+i0.307)$ \\
\hline
solution FA84 \cite{SAID} &    $(0.400+i0.339)$    &    $(0.348+i0.296)$ \\
\hline
solution CV90 \cite{SAID} &    $(0.412+i0.328)$    &    $(0.356+i0.283)$ \\
\hline
solution KV90 \cite{SAID} &    $(0.418+i0.334)$    &    $(0.350+i0.279)$ \\
\hline
solution SM90 \cite{SAID} &    $(0.407+i0.303)$    &    $(0.379+i0.282)$ \\
\hline
solution FA93 \cite{SAID} &    $(0.403+i0.344)$    &    $(0.344+i0.293)$ \\
\hline
solution WI94 \cite{SAID} &    $(0.408+i0.339)$    &    $(0.348+i0.289)$ \\
\hline
solution SP95 \cite{SAID} &    $(0.399+i0.338)$    &    $(0.350+i0.296)$ \\
\hline\hline
\end{tabular}
\end{center}
\caption[table1]{
      The $\eta N$ S-wave scattering length within the framework of  any
      single resonance model, using $\pi N$ elastic T matrix at $\eta N$
      threshold     from     different     phase     shift      analyses
      \cite{Cut79,Hoe83,Ka84,SAID}  and  $\eta$  production  total cross
      section \cite{Bin73} from Eq.~(\ref{eq:XStot}) as the input.   The
      $\pi N$ T matrix is given in the first column, and resulting $\eta
      N$ S-wave scattering length in the second one.  }
\end{table}

\begin{table}[h]
\label{tab:2}
\begin{center}
\begin{tabular}{cccc}
\hline\hline
            &   $a_{\eta N}$  &    $T_{\pi \, \pi}$(1487 MeV) &
 $\sigma^{\rm tot}_{\pi^-p\rightarrow \eta n}(W_\eta)/p_\eta$         \\
            &      [fm]       &                                 &
 [$\mu$b/MeV]                                                         \\
\hline\hline
Bhalerao-Liu \cite{Bha85}  & $(0.27+i0.22)$ & $(0.38+i0.31)^*$ & 15.0 \\
                           & $(0.28+i0.19)$ & $(0.37+i0.25)^*$ & 13.4 \\
\hline
"modified"  Bhalerao-Liu \cite{Bat95}
                           & $(0.38+i0.31)$ & $(0.38+i0.31)^*$ & 21.2 \\
                           & $(0.44+i0.30)$ & $(0.37+i0.25)^*$ & 21.2 \\
\hline
Abaev and Nefkens \cite{Aba94}
                           & $(0.62+i0.30)$ & $(0.298+i0.145)$ & 20.2 \\
\hline
Our SR model \cite{Bat95}  &$(0.404+i0.343)$& $(0.345+i0.293)$ & 21.2 \\
\hline
Arima et al.  \cite{Ari92} & $(0.98+i0.37)$ &     --           & --   \\
\hline
Wilkin \cite{Wil93}        & $(0.55+i0.30)$ &     --           & --   \\
\hline
Our full model \cite{Bat95}&$(0.886+i0.274)$&     --           & --   \\
                           &$(0.876+i0.274)$&     --           & --   \\
\hline\hline
\end{tabular}
\\[0.1cm]
$^*$ data are read of the graphs
\end{center}
\caption[table2]{
      Results  of  other  analyses  for  the  $\eta N$ S-wave scattering
      length.  The scattering length is given in first column,  and  for
      single  resonance  models for which Eq.~(\ref{eq:aetan}) is valid,
      the  $\pi  N$  elastic  T  matrix  at  $\eta  N$   threshold   and
      corresponding  $\eta$ production total cross section near threshold
      are given in last two columns.  }
\end{table}

\begin{table}[h]
\label{tab:3}
\begin{center}
\begin{tabular}{cccr}
\hline\hline
     & $a_{\eta N}$ [fm] & $T_{\pi \,\pi }$(1487 MeV) & $R$\ \ \ \\
\hline\hline
Bhalerao-Liu \cite{Bha85}
        &  $(0.27+i0.22)$  &  $(0.38+i0.31)^*$    & $-0.001$ \\
        &  $(0.28+i0.19)$  &  $(0.37+i0.25)^*$    & $ 0.002$ \\
\hline
Abaev and Nefkens \cite{Aba94}
        &  $(0.62+i0.30)$  & $(0.298+i0.145)$     & $ 0.002$ \\
\hline
Our SR model \cite{Bat95}
        & $(0.404+i0.343)$ & $(0.345+i0.293)$     & $ 0.000$ \\
\hline
\hline
Arima et al.  \cite{Ari92}
        &  $(0.98+i0.37)$  &  $(0.47+i0.37)^*$    & $-0.316$ \\
\hline
Our SRBG \cite{Bat95}
        & $(0.691+i0.174)$ & $(0.381+i0.392)$     & $-0.618$ \\
\hline
Our CSRBG \cite{Bat95}
        & $(0.968+i0.281)$ & $(0.413+i0.317)$     & $-0.391$ \\
\hline
Our full model \cite{Bat95}
        & $(0.886+i0.274)$ & $(0.373+i0.331)$     & $-0.453$ \\
        & $(0.876+i0.274)$ & $(0.375+i0.330)$     & $-0.446$ \\
\hline\hline
\end{tabular}
\\[0.1cm]
$^*$ data are read of the graphs
\end{center}
\caption[table3]{
      The test of the single resonance character  of  different  models.
      The  $\eta  N$  S-wave scattering length is given in first column,
      and the $\pi N$ elastic T matrix at $\eta N$ threshold  in  second
      one.  The ratio defined in Eq.~(\ref{eq:SRR}) is given in the last
      column.  For the single resonance models this ratio must be  close
      to  zero.   Deviations  from zero can be used as a measure for the
      importance of the background terms and/or higher resonances.}
\end{table}
\bigskip

\clearpage
\section*{Figure Caption}
      {\bf Fig.~1.} $\eta  N$  S-wave scattering length.\\
      Open  circles  represent  the  $\eta  N$  S-wave scattering length
      values obtained on the basis of  $\pi  N$  elastic  T-matrices  of
      references  \cite{Cut79,Hoe83,Ka84,SAID} and the $\eta$ production
      total cross section given in  Eq.~(\ref{eq:XStot}).   Full  normal
      and   inverse   triangles  represent  the  original  and  modified
      scattering length values  of  Ref.   \cite{Bha85},  and  the  full
      square  gives  the  value  given  by Ref.  \cite{Aba94}.  The full
      circles represent the value  of  Ref.~\cite{Wil93}.   Open  square
      represent  the  value  obtained  in  our  full  model \cite{Bat94}
      restricted to the single resonance only and  an  open  normal  and
      inverse triangles represent the values of our full, multiresonance
      model with three and four resonances in the P$_{11}$ partial wave,
      respectively.    The   full   star   represents   the   value   of
      Ref.~\cite{Ari92}.

\end{document}